\pdfoutput=1
\RequirePackage{ifpdf}
\documentclass[12pt,a4paper]{article}
\usepackage[utf8]{inputenc}
\usepackage[english]{babel}
\usepackage{amsmath}
\usepackage{jinstpub}
\usepackage{lineno}
\usepackage[section]{placeins}
\usepackage{float}
\usepackage{soul}
\usepackage{mathtools}
\usepackage{graphicx}
\usepackage[colorinlistoftodos]{todonotes}
\usepackage{multirow}
\usepackage{booktabs}
\usepackage{bm}
\usepackage{subcaption}
\usepackage{fancyhdr}
\collaboration{MicroBooNE Collaboration}

\author[g]{R.~Acciarri}
\author[bb]{C.~Adams}
\author[h]{R.~An}
\author[c]{J.~Anthony}
\author[y]{J.~Asaadi}
\author[a]{M.~Auger}
\author[g]{L.~Bagby}
\author[bb]{S.~Balasubramanian}
\author[g]{B.~Baller}
\author[n]{C.~Barnes}
% ...

\affiliation[a]{Universit{\"a}t Bern, Bern CH-3012, Switzerland}
% ...
\affiliation[c]{University of Cambridge, Cambridge CB3 0HE, United Kingdom}
% ...
\affiliation[g]{Fermi National Accelerator Laboratory (FNAL), Batavia, IL 60510, USA}
\affiliation[h]{Illinois Institute of Technology (IIT), Chicago, IL 60616, USA}
% ...
\affiliation[n]{University of Michigan, Ann Arbor, MI, 48109, USA}
% ...
\affiliation[y]{University of Texas, Arlington, TX, 76019, USA}
% ...
\affiliation[bb]{Yale University, New Haven, CT, 06520, USA}

\title{Michel Electron Reconstruction Using Cosmic-Ray Data from the MicroBooNE LArTPC}

\abstract{The MicroBooNE liquid argon time projection chamber (LArTPC) has been taking data at Fermilab since 2015 collecting, in addition to neutrino beam, cosmic-ray muons. Results are presented on the reconstruction of Michel electrons produced by the decay at rest of cosmic-ray muons. Michel electrons are abundantly produced in the TPC, and given their well known energy spectrum can be used to study MicroBooNE's detector response to low-energy electrons (electrons with energies up to \textasciitilde $50$~MeV). We describe the fully-automated algorithm developed to reconstruct Michel electrons, with which a sample of \textasciitilde 14,000 Michel electron candidates is obtained. Most of this article is dedicated to studying the impact of radiative photons produced by Michel electrons on the accuracy and resolution of their energy measurement. In this energy range, ionization and bremsstrahlung photon production contribute similarly to electron energy loss in argon, leading to a complex electron topology in the TPC. By profiling the performance of the reconstruction algorithm on simulation we show that the ability to identify and include energy deposited by radiative photons leads to a significant improvement in the energy measurement of low-energy electrons. The fractional energy resolution we measure improves from over 30\% to \textasciitilde 20\% when we attempt to include radiative photons in the reconstruction. These studies are relevant to a large number of analyses which aim to study neutrinos by measuring electrons produced by $\nu_e$ interactions over a broad energy range.}

\keywords{Michel electrons, LArTPC, MicroBooNE}

\begin{document}
\maketitle
\section{Introduction}
 
\paragraph{} The study of neutrino oscillations offers an exciting opportunity to expand our knowledge of fundamental particle physics, and offers a clue to explore the existence of new physics. Neutrinos are studied by measuring the products of their interactions with matter. Accurately measuring electrons from $\nu_e$ interactions is necessary for upcoming research programs investigating non-standard neutrino oscillations, lepton-sector CP violation, and the neutrino mass-hierarchy~\cite{bib:SBN,bib:DUNEcdrPhysics}. Additionally, neutrinos produced in galactic supernovae bursts (SNBs) are expected to produce electrons in the $5$--$50$~MeV energy range~\cite{bib:1987a_IMB,bib:1987a_Kamiokande}. The ability to obtain a precise energy response for low-energy electrons is therefore essential for neutrino oscillation and SNB measurements. Therefore, this work is relevant for the study of electrons produced by $\nu_e$ interactions from supernova bursts, low-energy $\nu_e$ interactions from accelerator sources, and low energy $\gamma$ electromagnetic (EM) showers produced by neutral current interactions in a wide range of neutrino energies.

\par The LArTPC detector technology has been chosen to pursue these programs in the U.S.A. at Fermilab through a number of short- and long-term projects. The Short-Baseline Neutrino program~\cite{bib:SBN}, aimed at studying possible non-standard neutrino oscillations, consists of a suite of three LArTPCs (SBND, MicroBooNE, and ICARUS). MicroBooNE is currently in operation, and will be followed by the full three detector program starting in 2019. DUNE (Deep Underground Neutrino Experiment) is a long-baseline neutrino experiment planning to build multi-kiloton-scale LArTPC detectors at the Long Baseline Neutrino Facility in South Dakota~\cite{bib:DUNEcdrPhysics,bib:DUNEcdr} to make precise neutrino oscillation measurements with the aim of investigating CP violation in the lepton-sector, the neutrino mass hierarchy, as well as studying SNB neutrinos. Furthermore, several detector prototypes and technology demonstrators have taken valuable data (LArIAT~\cite{bib:lariat}, LAPD~\cite{bib:LAPD}, CAPTAIN~\cite{bib:captain}, LBNE 35 ton prototype~\cite{bib:35ton}) or are under construction (protoDUNE single- and dual-phase detectors).

\par Experimental validation of the LArTPC technology's ability to achieve the electron energy resolution necessary for the rich physics program envisioned is incomplete and requires further study and validation. The ArgoNeuT~\cite{bib:argoneut_intro} and ICARUS T600~\cite{bib:icarus_intro} detectors have made a series of measurements useful in addressing questions regarding the ability of LArTPCs to provide precise calorimetric energy reconstruction of EM activity. Studies of recombination~\cite{bib:icarus_recomb,bib:argoneutRecomb} and electron absorption~\cite{bib:icarus_lifetime} address the ability to recover information about the energy deposited in the TPC volume. Measurements of $\gamma$ EM showers produced by $\pi^0$ decay~\cite{bib:icarus_pi0,bib:argoneut_pi0}, as well as the calorimetric separation of electron and $\gamma$ EM showers~\cite{bib:argoneut_dedx}, provide information specific to the reconstruction of EM shower kinematic properties. In the low-energy range, ICARUS has produced a measurement of the $\mu$ decay spectrum~\cite{bib:icarus_michel}. However, detailed investigations of the impact of the topology of EM-shower propagation in LArTPCs on energy reconstruction are currently lacking.

\par MicroBooNE~\cite{bib:microboone_detectorpaper} is a 170 metric ton LArTPC currently operating at Fermilab in the Booster Neutrino Beamline. The detector has been operational since summer 2015 and began collecting data with a stable neutrino beam in October of 2015. Data from the MicroBooNE detector can help us better understand electron reconstruction in a LArTPC detector.

\par
\begin{figure}[H]
\centering
\includegraphics[width=0.9\textwidth]{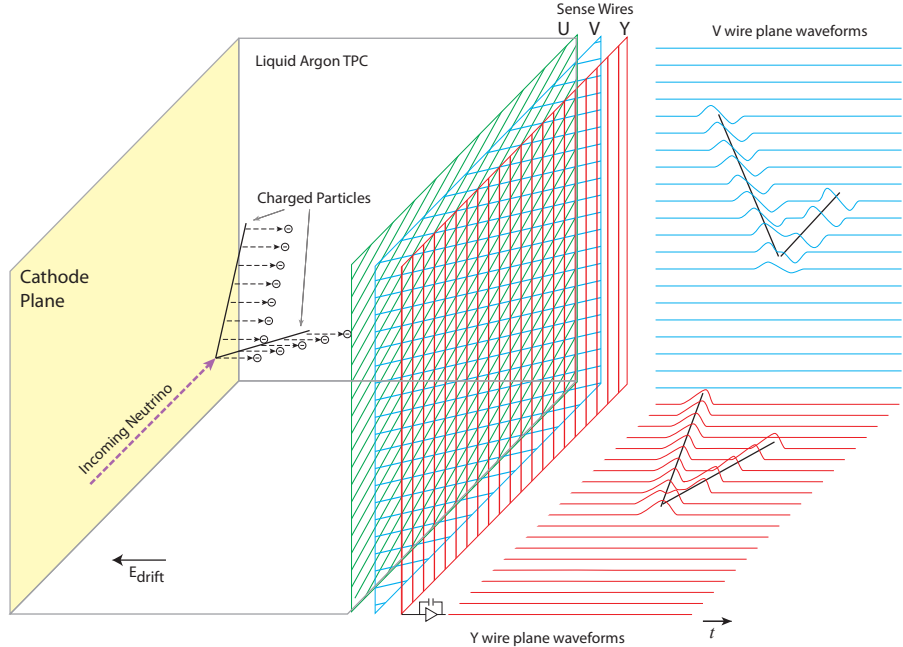}
\caption{Working principle of the MicroBooNE TPC. A constant electric field applied to the cathode (left-hand side) produces a uniform drift-field of 273 V/cm in the drift-coordinate direction. Charged particles traversing the TPC, such as a cosmic-ray muon, will ionize the argon, depositing a trail of ionized electrons, which will then begin to drift towards the wire planes under the effect of the electric field. Charge drifts at \textasciitilde0.11 cm/$\mu$s~\cite{bib:larprop}.}
\label{fig:tpcsketch}
\end{figure}

 The MicroBooNE TPC is a 10.37-m (beam direction) $\times$ 2.56-m (drift direction) $\times$ 2.32-m (vertical direction) parallelepiped placed within a cylindrical cryostat. The TPC encompasses approximately 90 metric tons of argon, which make up the detector active volume. Figure~\ref{fig:tpcsketch} shows a schematic of MicroBooNE's TPC working principle. Charged particles traversing the detector volume ionize the argon, leaving behind a trail of electrons along the particle's original trajectory. A uniform electric field causes electrons to drift towards three closely-spaced parallel wire planes. A total of 8,256 wires record signals from the TPC.  The first two planes that a drifting electron encounters respond with bipolar “induction plane” signals, while the third plane responds with a unipolar “collection plane” signal.  The anode wires in the three planes are oriented at +60, -60, and 0 degrees with respect to vertical, which combined with the drift time measurement allows for three-dimensional (3D) reconstruction of the drift electron production point via three independent drift-time measurements. Wires are separated by $3$~mm, and TPC data is digitized at $2$~MHz, allowing the recording of highly detailed images of interactions in the TPC. The high signal-to-noise ratio of MicroBooNE's electronics~\cite{bib:microboone_noise} enables the recording of precise calorimetric information. In this analysis we only use the collection plane, which provides a two-dimensional (2D), top-down projection of the charge deposited in the TPC.

\par This work provides an experimental measurement of the response of low-energy electrons  using the MicroBooNE LArTPC. In order to do so, we use \textit{Michel electrons}, electrons produced by the decay-at-rest of cosmic-ray muons that come to a stop in the MicroBooNE TPC. Michel electrons have a well characterized energy spectrum~\cite{bib:michelNevis} with energies up to \textasciitilde $50$~MeV. We can use these electrons to study the response to neutrino interactions which produce EM activity in the same energy range.

\par We have developed a fully-automated set of algorithms to reconstruct such EM showers in our LArTPC. We describe the features and challenges of this reconstruction in the following sections.  We show that Michel electrons manifest themselves with a complex topology caused by the stochastic release of bremsstrahlung photons~\cite{bib:bremreview}, which can disperse significantly and carry away a considerable fraction of the electron's energy. Due to the 14 cm radiation length in argon~\cite{bib:radlenPDG} (corresponding to 47 anode-plane wires), and the relatively slow drift-velocity in the MicroBooNE TPC, collecting the total charge deposited by Michel electrons is made more challenging by the need to identify the charge of bremsstrahlung photons. We show that collecting radiative photons is essential in order to obtain good energy resolution for low energy-electrons.

\par  In this article we report on the current status of reconstructing Michel electrons in MicroBooNE. We present results from a two-dimensional approach, which employs only information from MicroBooNE's collection plane. Section~\ref{sec:edep} describes the general signature of Michel electrons in a LArTPC, presenting the unique topological features of low-energy electrons and the challenges encountered when trying to reconstruct their energy. Section~\ref{sec:recoalg} describes the algorithm developed to tag and reconstruct Michel electrons from cosmic-ray muons. The method by which we reconstruct the energy using calorimetric information appears in Sec.~\ref{sec:ereco}. Section~\ref{sec:eres} presents results of a Monte Carlo (MC) study on the energy reconstruction for Michel electrons performed by using the algorithm described in Sec.~\ref{sec:recoalg}. Section~\ref{sec:espectrum} presents the reconstructed energy spectrum obtained for data, and Section~\ref{sec:conclusions} offers some concluding remarks.

\section{Low-Energy Electrons in Liquid Argon}
\label{sec:edep}

\subsection{Energy Deposition in Liquid Argon}

\paragraph{}Michel electrons are detected by the MicroBooNE TPC through ionization energy loss by the primary electron originating from the muon decay, and the secondary electrons produced by bremsstrahlung photons that undergo Compton scattering or pair-production. To identify Michel electrons and to reconstruct their energy it is important to understand exactly what signature an electron will leave as it traverses the argon. In this section we explore the topology of low-energy electrons in a LArTPC, presenting a review of the well-known physics processes which govern the propagation of electrons through argon. Ultimately we are interested in determining how this will impact electron energy reconstruction. Simulation results presented in this section are obtained through a Geant4~\cite{bib:geant4} simulation of particle propagation in the MicroBooNE detector. Useful properties of Michel electrons and LArTPCs referred to throughout this section are listed in Table~\ref{tab:tab}.

\begin{table}[tb]
\begin{center}
\caption{Table summarizing useful properties of the MicroBooNE TPC and of energy deposition in LAr.}
\label{tab:tab}
\begin{tabular}{ | l | l |}
\hline
{\bf Quantity} & {\bf Value} \\ \hline
$\mu^+$ mean lifetime & 2.2 $\mu$s~\cite{bib:pdg} \\ \hline
Michel electron energy range & 0-60 MeV \\ \hline
Photon absorption length in LAr & \textasciitilde 27 cm~\cite{bib:NIST_XCOM} \\ \hline
Radiation length in LAr &  14 cm~\cite{bib:radlenPDG} \\ \hline
Electron critical energy in LAr & 45 MeV~\cite{bib:NIST_ESTAR} \\ \hline
dE/dx for ionization for electrons in LAr & 1.9-2.7 MeV/cm~\cite{bib:NIST_ESTAR} \\ \hline
MicroBooNE drift velocity & 0.11 cm / $\mu$s~\cite{bib:larprop} \\ \hline
MicroBooNE wire spacing & 0.3 cm~\cite{bib:microboone_detectorpaper} \\ \hline
\end{tabular}
\end{center}
\end{table}

\par
\begin{figure}[H]
\centering
\includegraphics[width=0.7\textwidth]{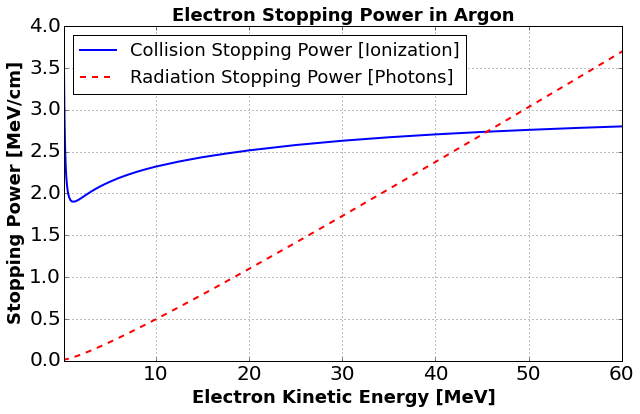}
\caption{Energy loss per unit distance (in MeV/cm) for electrons traveling through liquid argon. Collision (solid blue line) and radiation (dashed red line) stopping power are shown separately. Data shown here are taken from the NIST ESTAR database~\cite{bib:NIST_ESTAR} and converted to MeV/cm by accounting for a density of liquid argon of 1.38 g/${\rm cm^3}$.}
\label{fig:energyloss_NIST}
\end{figure}

 Electrons traveling in a medium lose energy either by ionizing and exciting the atoms which make up the material (collision stopping power)  or by producing photons from a bremsstrahlung interaction (radiative stopping power). At energies well beyond the critical energy, radiative loss greatly exceeds ionization loss, causing the production of EM showers in which a cascade of electrons and photons produces a dense cloud of charge deposition pointing in the direction of the original electron's momentum. These showers are generated if the energy of the radiative photons is high enough to produce electron-positron pairs, which can, in turn, emit other bremsstrahlung photons. A plot showing the relative contribution to energy-loss for electrons up to 60 MeV in energy in liquid argon is shown in figure~\ref{fig:energyloss_NIST}. One can notice that the relative contribution of ionization and bremsstrahlung varies significantly across the energy range of Michel electrons. Moreover, the energy range of interest for Michel electrons includes the critical energy at which contributions from bremsstrahlung and ionization are equal, and below which the shower extinguishes.

\par
\begin{figure}[H]
\centering
\includegraphics[width=1.0\textwidth]{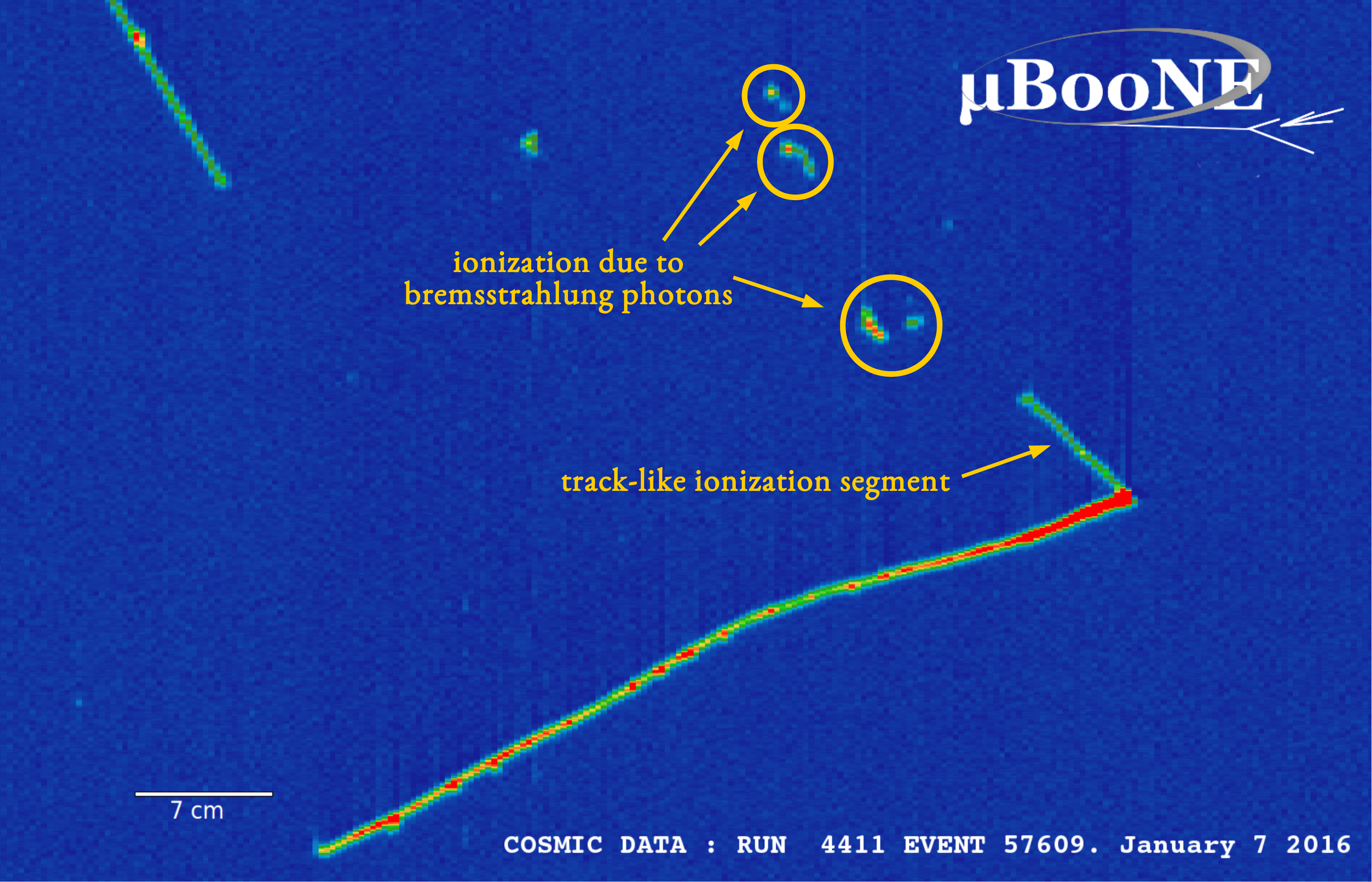}
\caption{Candidate Michel electron event from cosmic-ray data, which produces a track-like segment of deposited charge due to ionization as well as several radiative photons. This image shows raw data from MicroBooNE's collection-plane view. Signals from each wire fall along vertical lines, with wire number increasing towards the right. The vertical axis shows the charge collected at different times, with time increasing in the positive y direction. The scale bar applies to both the horizontal and vertical coordinates. The color scale denotes the charge collected by the sense wires, with higher charge collected denoted in red.}
\label{fig:ionization_vs_radiation_evd}
\end{figure}

 Energy deposition via ionization is continuous in the TPC, producing track-like segments. Photons, however, will not deposit energy immediately in the TPC but rather will travel undisturbed until they either Compton scatter or produce an $e^+ / e^-$ pair. The significant distances traveled by photons before depositing energy in the TPC makes their signature qualitatively different from that of the Michel electron's primary ionization trail. Figure~\ref{fig:ionization_vs_radiation_evd} shows an example candidate Michel electron from data where energy deposited via ionization and radiative photons are clearly distinguishable from each other.

\par
\begin{figure}[H]
\centering
\includegraphics[width=0.8\textwidth]{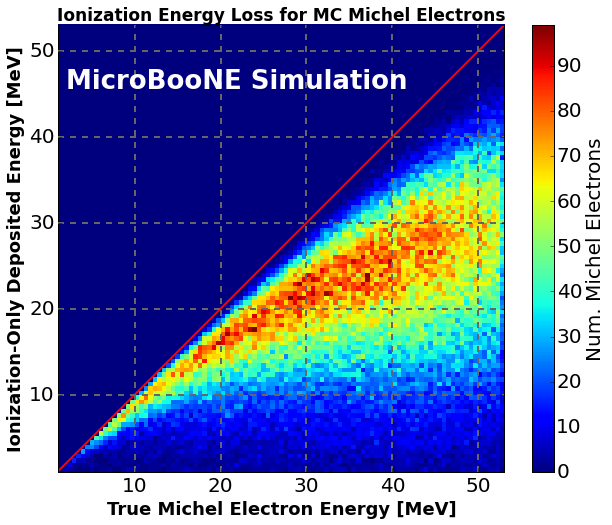}
\caption{Comparison of the simulated energy deposited only via primary ionization as a function of the total Michel electron energy. In the scatter plot the true electron energy and the energy deposited solely via primary ionization are plotted on the x and y axes, respectively.}
\label{fig:Egammafrac}
\end{figure}

 Bremsstrahlung-photon production~\cite{bib:bremreview} has a significant impact on the topology of Michel electrons in liquid argon. Photons are created via bremsstrahlung stochastically: case-by-case, the number of photons produced by an electron as it travels in the detector and the energy distribution of these photons will vary significantly. The impact this has on Michel electrons can be gauged by studying figure~\ref{fig:Egammafrac} where we plot the simulated energy deposited via ionization by each Michel electron as a function of the true total Michel electron energy. By studying this distribution one notices that the fraction of energy lost to photons varies significantly depending on the electron's initial energy. Lower energy electrons tend to deposit most of their energy as ionization and only a small contribution to the total energy deposition comes from bremsstrahlung photons. The higher the electron energy, the larger the contribution from photons. This conclusion is consistent with what is shown in figure~\ref{fig:energyloss_NIST}. Additionally, one sees that the variation in the fractional energy lost to radiative photons varies considerably case-by-case due to the stochastic nature of bremsstrahlung production.

\par
\begin{figure}[H]
\centering
\includegraphics[width=0.55\textwidth]{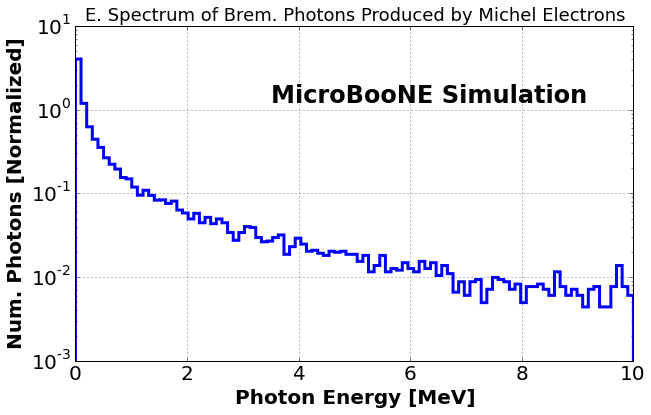}
\caption{Energy spectrum of bremsstrahlung photons produced by a sample of Michel electrons from MicroBooNE simulation. The distribution highlights the production of many low-energy photons.}
\label{fig:photon_espectrum}
\end{figure}

 Figure~\ref{fig:photon_espectrum} shows the energy distribution for photons produced via electron bremsstrahlung by Michel electrons. The fact that many low energy photons are produced can complicate the attempt to collect the full energy released by Michel electrons, as very low energy photons may produce signals which are below threshold in the TPC.

\par
\begin{figure}[H]
\centering
\includegraphics[width=0.7\textwidth]{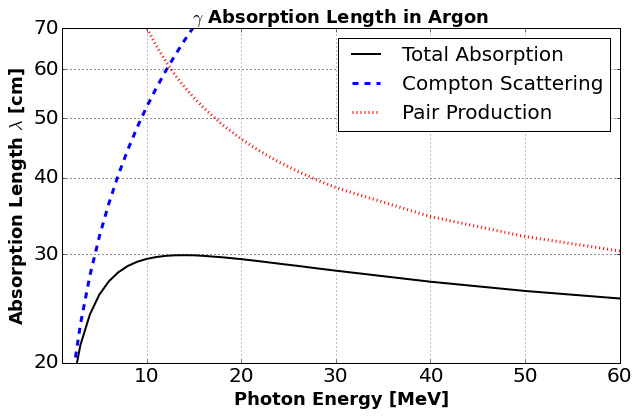}
\caption{Absorption length $\lambda$ [cm] in argon as a function of the photon's energy in MeV. Data shown here are taken from the NIST XCOM database~\cite{bib:NIST_XCOM}. The total absorption curve (black, solid) includes contributions from pair production (red, dotted), Compton scattering (blue, dashed), and additional processes which are subdominant in this energy range. Slightly above 10 MeV the contribution from pair production takes over that from Compton scattering as the dominant interaction process. Both effects combined lead to an approximately constant attenuation length of $20$--$30$~cm across most of the energy range shown.}
\label{fig:absorptionlength_NIST}
\end{figure}

 The energy-dependent absorption length for photons in liquid argon is plotted in figure~\ref{fig:absorptionlength_NIST}. In the energy range of interest for Michel electrons both Compton scattering and pair-production play a significant role.

\par
\begin{figure}[H]
\centering
\includegraphics[width=0.6\textwidth]{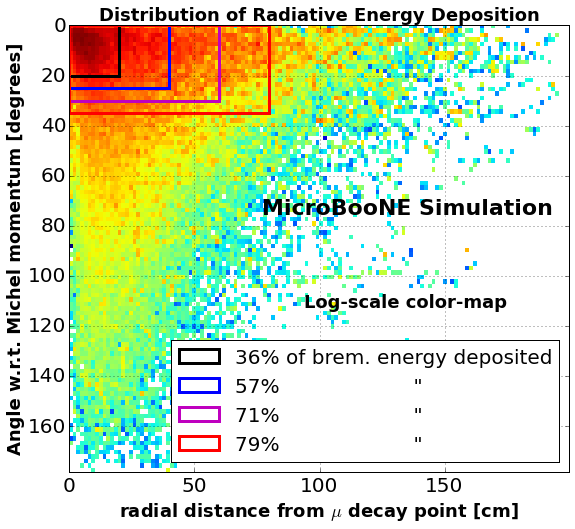}
\caption{Distribution of energy deposited by radiative photons with respect to the distance from the decay point of the muon and the direction with respect to the original momentum of the electron. Colored boxes are labeled by the fraction of total deposited energy within their bounds.}
\label{fig:photondistrib}
\end{figure}

 We next produce a  profile of the spatial distribution of charge deposited via radiative photons using information from a simulation of Michel electrons propagating in the MicroBooNE TPC. Figure~\ref{fig:photondistrib} shows the distribution of energy deposited by photons as a function of the radial distance from the muon decay point and the angular separation with respect to the electron's initial momentum direction. Colored boxes on the plot are labeled by the fraction of total energy deposited by secondary bremsstrahlung photons within their bounds. The large distances over which energy is deposited can pose a challenge when attempting to collect the total visible energy released by Michel electrons. The fact that the spatial profile of charge deposited is characterized by a few low energy photons, rather than a fully developed EM shower, makes tagging the individual photons more challenging. This is particularly true for a surface detector such as MicroBooNE, where a large cosmic-ray flux creates showers uncorrelated with the Michel electron. This problem should be mitigated by the cosmic-ray tagger detector (CRT) that we have recently installed in MicroBooNE~\cite{bib:CRT}.

\subsection{Muon Decay and Capture in Liquid Argon}

\paragraph{}An important effect is the distortion of the Michel-electron energy spectrum due to decay from muons bound to atoms in a medium. Negatively charged muons that come to rest in a medium can be captured by atoms of that medium. These muons travel in an orbit close to the nucleus and can be absorbed by the nucleus before decaying. If the muon does decay its orbital velocity will cause the energy spectrum of the electron to broaden. Calculations of the observable Michel energy spectrum due to bound-muon decay have been performed~\cite{bib:boundmudecay}. These effects are accounted for in the MicroBooNE Monte Carlo, which employs CORSIKA~\cite{bib:corsika} for generating the flux of cosmic ray particles entering the detector, and Geant4~\cite{bib:geant4} for the propagation of particles through the detector volume.

\par
\begin{figure}[H]
\centering
\includegraphics[width=0.7\textwidth]{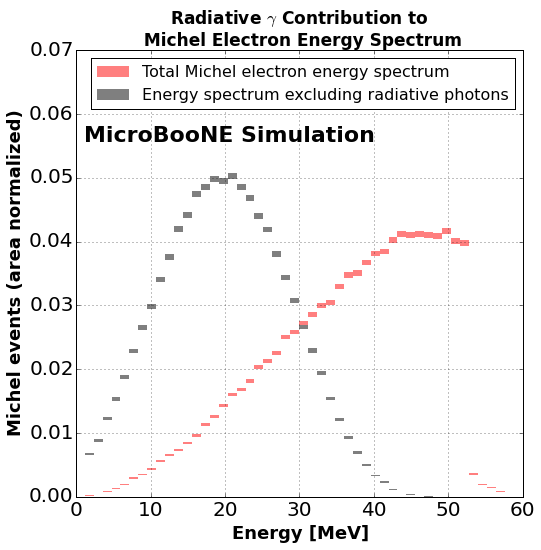}
\caption{Energy spectrum for Michel electrons from a Monte Carlo sample of generated cosmic-ray events in MicroBooNE. The simulation takes into account distortions to the Michel electron spectrum due to the different decay probabilities and energies for $\mu^+$ and $\mu^-$. In red we show the true Michel electron energy spectrum, and in black the ionization-only deposited energy, excluding energy loss to radiative photons. Both distributions are area-normalized.}
\label{fig:Espectrum_MC_all_norad}
\end{figure}

 We conclude this discussion on the impact of different physics effects on the Michel electron energy spectrum expected for MicroBooNE by showing what the Michel electron energy spectrum looks like both including and excluding all energy deposited by radiative photons in the TPC. Figure~\ref{fig:Espectrum_MC_all_norad} shows the Michel energy spectrum distribution in red and the spectrum if all energy deposited via radiative photons is excluded (``ionization-only'' energy) in black. The spectrum without radiative photons is significantly shifted to lower energies. Furthermore, it appears to peak at around 20 MeV. This is due to the fact that higher energy electrons lose more energy to radiative photons than lower-energy ones, sculpting the true distribution to exhibit a peak. The flattening of the Michel-electron total-energy spectrum at around 40 MeV, with the presence of a tail extending to 60 MeV, is a consequence of the decay of bound $\mu^-$.

\section{Michel-Electron Reconstruction}
\label{sec:recoalg}

\paragraph{}In this section we present the method by which we identify Michel electrons in the TPC. Michel electrons deposit charge over a region which spans tens of centimeters. With a drift velocity of 1.1 cm / 10 $\mu$s the arrival time of the drifting electron is spread over hundreds of $\mu$s. This timescale is much larger than the 2.2 $\mu$s muon lifetime and indicates that we cannot isolate and tag charge based on its arrival time on the sense-wires. Instead, we take advantage of the high-resolution topological and calorimetric information provided by the MicroBooNE TPC to search for signatures in the detector that are characteristic of a stopping muon producing a Michel electron.

\par
\begin{figure}[h]
\centering
\includegraphics[width=0.9\textwidth]{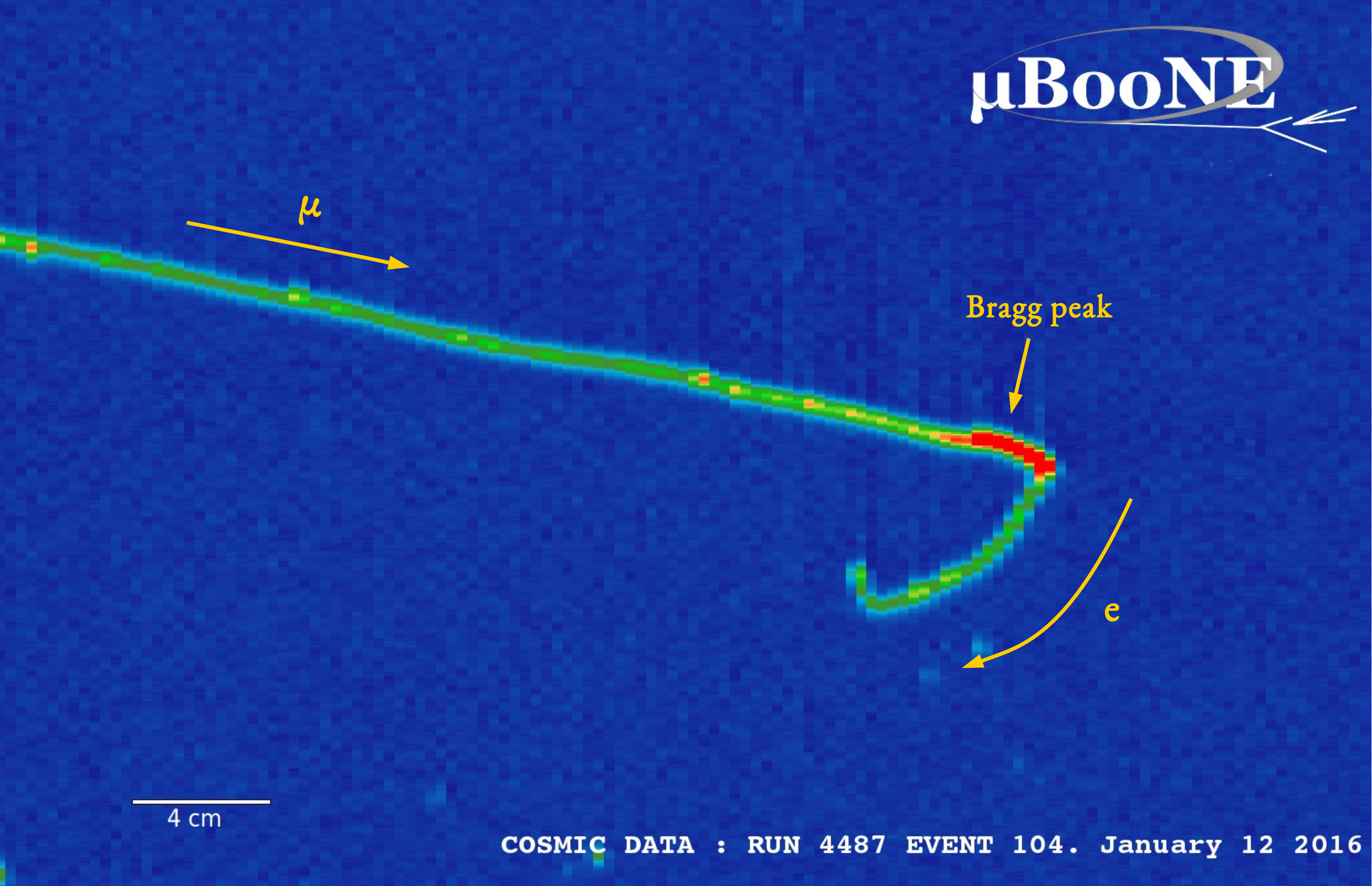}
\caption{Event display showing raw data from a small region of the TPC volume from the collection plane where a candidate Michel electron was identified. The x axis shows the data along the beam direction (increasing wire-number to the right) and the y axis the drift-coordinate (increasing drift-time moving upwards). The scale bar applies to both the horizontal and vertical coordinates. The color coding denotes the amount of collected charge on each wire per time tick. In this display the muon candidate coming to a stop can be identified by the significant increase in charge deposited per unit distance at the Bragg peak. The outgoing Michel electron is distinguishable
as the short track originating at the muon’s stopping point and depositing charge for several centimeters in a different direction from that of the muon.}
\label{fig:evd2}
\end{figure}

 Michel electrons from MicroBooNE cosmic data (data taken with the detector using a strobe trigger, in off-beam mode) are identified using an automated reconstruction chain of algorithms that process the raw data into reconstructed quantities, which are then fed to pattern-recognition algorithms. As muons come to a stop, energy deposition along the particle's trajectory shows a characteristic rise, called the ~\textit{Bragg peak}. Most Michel electrons produced by the decay will propagate in a direction different than that of the stopping muon. We use information about the position and density of the charge deposited within the TPC to identify the Bragg peak and the kink between the Michel electron and the stopping muon. An event display of a well-identifiable Michel electron candidate can be seen in figure~\ref{fig:evd2}.

\par The reconstruction procedure begins with an initial signal processing stage, in which the raw digitized waveform from each wire is deconvolved to account for the detector’s field response and electronics shaping time. A software noise filter that suppresses frequencies associated with noise features produced within the TPC is also applied at this stage. This leads to an estimate of the drifted charge extracted in units of electrons ($e^-_{\rm reco}$) as a function of time.  A pulse-finding algorithm is then run on the processed waveforms to identify and measure the charge delivered to the TPC wires. This step produces reconstructed hits, which carry information on the wire and time of arrival of charge, as well as an uncalibrated measure of the energy deposited in each hit, in units of $e^-_{\rm reco}$. Using a sample of crossing cosmic-ray muons the variation in detector response across the collection plane is measured to be of order 6\%. Hit wire and time information on the collection plane represents a 2D top-down projection of charge deposited within the TPC.

\par Once two dimensional hits have been reconstructed, a pattern-recognition clustering algorithm targeted to identify cosmic muons is run to associate hits that belong to the same particle or interaction. The clustering algorithm takes into account the relative spatial positioning of hits using both wire number and time information, in order to decide how hits should be clustered together. Hits associated to each localized cluster are then sorted based on their spatial orientation, in order to obtain a profile of the muon's charge deposition as it travels through the detector, which is smoothed by using a truncated mean algorithm. Given a spatially sorted list of 2D hits, the \emph{truncated mean charge} of each hit is calculated by taking the charge of neighboring hits, removing the upper and lower tails of the distribution of charge for this subset of hits, and then computing the mean of the values remaining. For this work we include 8 hits on each side, and truncate the highest and lowest 25\% of values. \par A profile of the muon's local linearity along the cluster is also calculated. The \emph{local linearity} is a measure of the correlation of wire and time coordinates in a neighborhood of each reconstructed 2D hit. For each 2D reconstructed hit, the local linearity is given by the covariance of wire and time coordinates calculated using the hit in question, and the five hits preceding and following it in the sorted muon cluster, divided by the product of the standard deviation of wire coordinates and that of time coordinates. A local linearity of one indicates perfect correlation for the selected hits, and thus a straight line-segment.
\par
\begin{figure}[bt]
\centering
\includegraphics[width=1.0\textwidth]{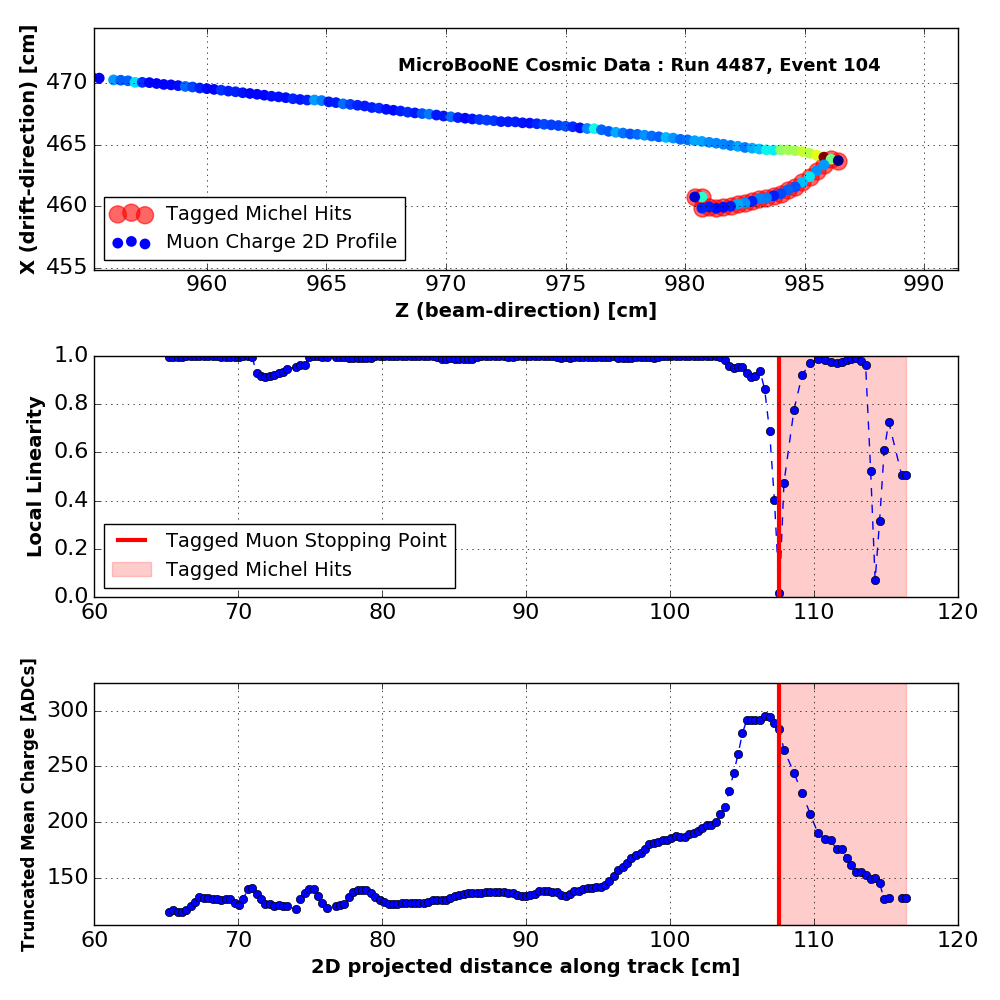}
\caption{Event display showing hits associated with a tagged Michel electron candidate (top). The middle and bottom panels show the local linearity and truncated mean charge profiles calculated for the muon plus Michel electron cluster. The red vertical line indicates the location of the identified stopping point for the muon, and the region shaded in red represents the set of hits along the cluster identified as belonging to the candidate Michel electron. In the top panel, the 2D event display shows information in time vs. wire coordinates. Each hit is represented as a circle, with the color indicating the charge measured. Blue indicates lower charge. Hits that are outlined in red represent those associated with the tagged Michel electron.}
\label{fig:clusterprofile}
\end{figure}

 Figure~\ref{fig:clusterprofile} shows reconstructed hits for the same event in figure~\ref{fig:evd2} with the calculated truncated mean and local linearity drawn in the bottom two panels of the image.

\subsection{Bragg Peak and Spatial Kink Identification} 
\paragraph{}As a muon comes to rest in the TPC and the density of deposited charge increases, the charge collected on each hit will increase as well. This will lead to a characteristic charge profile in which a Bragg peak from the stopping muon can be identified. The hit with the largest truncated mean charge is identified as the approximate location of the Bragg peak. Subsequently we measure the amount of charge deposited within the Bragg peak by integrating the truncated mean charge values in the last few cm up to the identified stopping point compared to that measured in the minimally ionizing particle (MIP) region of the muon to separate through-going from stopping muons. Finally, the hit with the largest charge within \textasciitilde 5 cm of the truncated mean charge maximum is reconstructed to be the 2D muon stopping location.
\par In a similar fashion we scan the muon track searching for a significant kink in the 2D topology of the projected charge. The average local linearity is calculated for the five hits surrounding the identified muon stopping point location, and we require this value to be smaller than 0.8. If both a clear Bragg peak and a significant spatial kink are found and all hits sorted after the reconstructed muon stopping point the reconstructed muon stopping point are tagged as belonging to a candidate Michel electron.  In order to further enhance the sample purity, a sequence of cuts on the candidate muon topology, requiring that at least 70\% of hits have a measured local linearity greater than 0.9 and that the muon segment be at least 10 cm long, target the removal of muon tracks which are short and exhibit a large amount of scattering, which make the identification of Michel electrons more difficult.

\subsection{Collecting Charge from Radiative Photons}
\paragraph{}A proximity-based charge-clustering algorithm will unavoidably miss any energy deposited by radiative photons. In order to attempt to recover any photons, we extend the search for charge associated with the electron in the direction of the electron itself. We begin by measuring a momentum direction for the Michel electron by fitting the charge deposited via ionization to a straight line. We then search for any charge deposited within an 80 cm distance (\textasciitilde 3 photon absorption lengths) of the tagged muon stopping point and within a \textasciitilde 30 degree opening angle with respect to the reconstructed electron direction. These cuts are applied to information on the collection plane, which is a 2D projection of 3D charge deposition points. From a simulation study similar to that performed in figure~\ref{fig:photondistrib} but in 2D, we expect 80\% of all energy deposited by bremsstrahlung photons to be contained within this cut range. Charge identified within this cut region is split into individual reconstructed photons by isolating contiguous clusters of hits. A minimum energy deposition of \textasciitilde 1 MeV is needed for a photon to be identified and included. Finally, we attempt to remove any tagged charge that may have been deposited by accidental cosmic ray muons. We do so by applying the following cuts: 
\begin{itemize}
\item any reconstructed photon with more than 20 hits (\textasciitilde 10 MeV) is excluded.
\item reconstructed photons with 5 to 20 hits are required to have a local linearity smaller than 0.8. If additional charge not tagged as belonging to the Michel electron is found in a projected 2 cm radius of any of the photon's hits, the photon is considered to be too close to an external cosmic-ray and thus discarded.
\end{itemize}
\par Without applying these cuts we find that the attempt to improve the energy measurement by including far-reaching radiative photons is hampered by the large amount of accidental cosmic activity crossing the event and mistakenly collected. For the case of Michel electrons the negative impact of incorrectly including charge deposited by other cosmic-ray interactions is particularly severe, given that just a few mis-identified hits can have a large impact on measuring the energy of a $< 50$ MeV electron. The first cut applied efficiently removes long muon tracks which cross the 80 cm, 30 degree cut region. The second removes smaller linear segments which may be due to activity in the proximity of cosmic-ray muons.

\begin{figure}[H]
\centering
\includegraphics[width=0.7\textwidth]{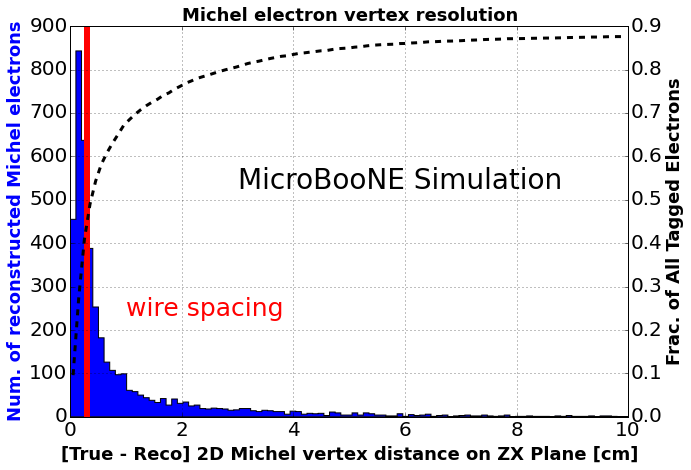}
\caption{Vertex resolution for reconstructed Michel electrons from a Monte Carlo cosmic sample. The histogram shows the 2D distance on the ZX-plane (wire- and drift-coordinates) of reconstructed Michel electrons to the nearest true Michel electron location. The vertical red line indicates the $3$~mm wire-spacing, and the dashed black curve quantifies the fraction of total reconstructed electrons with a 2D distance below a given value.}
\label{fig:vertexresolution}
\end{figure}

\paragraph{}The selection and reconstruction cuts applied above were tuned by visually inspecting their impact on data and simulation events and produce a pure sample of Michel electrons (80-90\% purity) with a reconstruction efficiency of approximately 2\%, which, given the large sample of cosmic data available, is sufficient to provide the statistics necessary for this study. The efficiency is quantified by measuring the number of reconstructed Michel electrons over the total number of electrons simulated in a Monte Carlo cosmic-ray event sample. The sample purity is estimated by comparing the wire and time vertex position of reconstructed vs. true Michel electrons. Figure~\ref{fig:vertexresolution} shows the result of this comparison. The fraction of all reconstructed events within a given distance from a true Michel electron vertex allows to estimate the sample purity. For example, 80\% of all reconstructed Michel electrons are within $3$~cm of a true muon stopping point. This plot also shows the good vertex resolution, peaking at \textasciitilde $2$~mm, below the wire-spacing. The low reconstruction efficiency is largely a consequence of the fact that we are only using 2D projected information to identify spatial and calorimetric features associated with the Michel electron topology. Backgrounds were studied by visually inspecting both data and MC events and were found to mostly consist of tagged EM activity from muons (delta-rays or bremsstrahlung photons) when such activity occurred close to the beginning or end of a muon track. A smaller subset of background events consisted of muon tracks which had undergone a large-angle scatter. Isolating background events in the simulation, we find their energy spectrum to be monotonically decreasing, with a low-end cutoff at \textasciitilde 10 MeV.

\section{Energy Reconstruction}
\label{sec:ereco}

\paragraph{}To determine the reconstructed energy we perform a calorimetric measurement by integrating the total charge tagged as being deposited by a Michel electron. A fixed calibration constant is applied to convert charge measured from reconstructed hits to MeV. This calibration accounts for the electronics of the collection-plane wires, the signal processing, as well as detector effects that convert the deposited energy into collected electrons on the wire planes. A correction factor used to convert from reconstructed electrons $e^-_{\rm reco}$ to true number of collected electrons $e^-$ on the wires is calculated using a sample of stopping muons, fitting the $dE/dx$ vs. residual range to values for liquid argon as tabulated by the PDG~\cite{bib:radlenPDG}. This leads to a correction factor
\begin{equation}
\label{eq:ecorr_calo}
\frac{e^-}{e^{-}_{\rm reco}} = 1.01.
\end{equation}

\par A number of physics processes are responsible for converting deposited energy to measurable electrons collected on the sense wires. We describe their impact and how we account for them in the following paragraphs.
\par
\begin{figure}[H]
  \centering
  \includegraphics[width=0.75\textwidth]{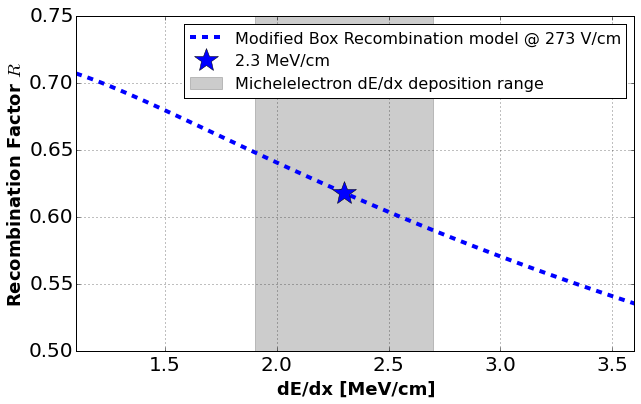}
  \caption{Recombination factor as a function of dE/dx computed using the Modified Box model as described by ArgoNeuT~\cite{bib:argoneutRecomb} for the value of the electric field strength in MicroBooNE of 273 V/cm.}
  \label{fig:recomb}
\end{figure}

 Recombination is the process by which ionized electrons are attracted by the positive ions produced along a particle's trajectory to re-form neutral argon atoms, which leads to a reduction of the number of drifting electrons. This physical process depends on dE/dx, the amount of energy deposited per unit distance, and the strength of the external electric field. In this analysis we model the effects of recombination according to the treatment and results obtained by the ArgoNeuT experiment~\cite{bib:argoneutRecomb}. The recombination factor $R$ is defined as
\begin{equation}
 \label{eq:recombdef}
dQ \times \frac{W_{{\rm ion}}}{e^-} = dE \times  R \left( \frac{dE}{dx}, E_{{\rm field}}\right).
\end{equation}
where $W_{\rm ion}$ denotes the work function for ionizing an argon atom (23.6 eV,~\cite{bib:Wion1,bib:Wion2}), $dQ$ the amount of ionization charge and $dE$ the amount of energy deposited over some distance. $dE/dx$ is the local energy deposition per unit distance and $E_{\rm field}$ the magnitude of the electric field, which is 273 V/cm for MicroBooNE. ArgoNeuT makes use of the Modified Box Model~\cite{bib:argoneutRecomb} to convert from charge to energy:
\begin{equation}
  \label{eq:modboxrecomb}
  \frac{dE}{dx} = \frac{ e^{\beta \times \frac{W_{ion}}{e^-} \frac{dQ}{dx}} - \alpha }{\beta}.
\end{equation}
Given equations~\ref{eq:recombdef} and~\ref{eq:modboxrecomb}, we can express the recombination factor $R$ as
\begin{equation}
  \label{eq:recombform}
  R = \frac{ \ln \left( \frac{dE}{dx} \times \beta + \alpha \right) }{ \frac{dE}{dx} \ \times \beta },
\end{equation}
with the constants $\alpha$ and $\beta$ given by
\begin{eqnarray}
  \alpha &=& 0.93 \pm 0.02, \\
  k_b &=& 0.212 \pm 0.002 \;\; {\rm \left[ \frac{g \times kV }{ MeV \times cm^3} \right] },\\
  \beta &=& \frac{k_b}{ \rho \; {\rm [g/cm^3]} \times E_{{\rm field}} \; {\rm [kV/cm]} } = 0.562 \;\; {\rm [cm / MeV]},
\end{eqnarray}
with $\rho$ denoting the density of liquid argon and $E_{\rm field}$ the strength of the electric field in the TPC. The values of $\alpha$ and $k_b$ come from ArgoNeuT's results.
\paragraph{} With this information we show in figure~\ref{fig:recomb} the value of the recombination factor as a function of dE/dx. We highlight in gray the range of dE/dx values of interest for Michel electrons, based on values from figure~\ref{fig:energyloss_NIST}.
\paragraph{}In this analysis we apply a constant recombination factor corresponding to a dE/dx value of 2.3 MeV/cm, which leads to a recombination factor of 0.62. The value of 2.3 MeV/cm is chosen as being half-way in the range 1.9 - 2.7 MeV/cm relevant to the collision stopping power for electrons with energies up to 60 MeV (see figure~\ref{fig:energyloss_NIST}). A Monte Carlo simulation of the effect of using a constant recombination factor leads to a 2\% spread in energy resolution for electrons in the energy range of Michel electrons.

\par Two additional detector effects which lead to position-dependent non-uniformities that can impact recorded signals are also considered. Impurities in the argon can absorb drifting electrons, thus quenching the signal being transported towards the sense wires. Because of its excellent filtration system, MicroBooNE has achieved very low concentrations of impurities, which minimize the loss of drifting electrons. Therefore no correction is applied to account for electron absorption by impurities. Space-charge effects, caused by the buildup of positive ions in the TPC, can lead to electric field distortions which in turn impact the calorimetry. A static space-charge simulation model, found to agree qualitatively with measurements from cosmic-ray data, shows that the impact on calorimetry throughout the detector volume leads to a smearing in the calorimetric energy measurement of less than 1\% and is thus considered negligible for the purpose of this analysis.
\par The energy of a Michel electron in MeV is given by multiplying the reconstructed charge ($e^-_{\rm reco}$) from hits associated with tagged Michel electrons by the calibration factor given by

\begin{eqnarray}
\label{eq:ADCtoE}
\frac{\rm E\,(MeV)}{e^-_{\rm reco}} &=&  1.01  \frac{e^-}{e^-_{\rm reco}} \times 23.6 \frac{\rm eV}{e^-} \times 10^{-6} \frac{\rm MeV}{\rm eV} \times \frac{1}{R} \\
&=& 3.85^{+ 0.21}_{-0.19} \times 10^{-5}. \nonumber
\end{eqnarray}

The uncertainty associated with the calorimetric energy conversion of Eq.~\ref{eq:ADCtoE} is meant to represent the systematic uncertainty in calculating a calorimetric energy conversion which can lead to a bias in our energy reconstruction. The evaluated uncertainty includes possible biases due to uncertainties in the ionization energy $W_{\rm ion}$, the treatment of charge quenching due to impurities, and the applied recombination model and correction. This corresponds to a $1^{+0.055}_{- 0.049}$ fractional systematic uncertainty.

\section{Effect of Radiative Photons on Energy Reconstruction: Monte Carlo Study}
\label{sec:eres}

\paragraph{}In this section we show the results of studying the impact of untagged radiative photons on the energy resolution by using a sample of simulated Michel electrons. We compare the reconstructed and true energy from the simulation under different scenarios.

\par We measure an energy-dependent fractional energy resolution by performing the following steps:
\begin{itemize}
	\item Plot the reconstructed energy as a function of the true energy and fit to the function $E_{\rm reco} = \alpha + \beta E_{\rm MC}$.
    \item Correct the energy scale by calculating a corrected energy $E_{\rm corrected} = \frac{E_{\rm reco} - \alpha}{\beta}$.
    \item Binning in true energy to measure the fractional resolution by fitting the distribution $[E_{\rm corrected} - E_{\rm MC}] / E_{\rm MC}$ to a Gaussian function. The $\sigma$ from the fit gives a measure of the fractional energy resolution.
\end{itemize}

\par The value $E_{\rm reco}$ is measured employing the calorimetric conversion defined in Sec.~\ref{sec:ereco}. The slope $\beta$ measures the bias between reconstructed and true energy introduced by the reconstruction algorithm, over and above what is already accounted for by the correction of Eq.~\ref{eq:ecorr_calo}. Both the energy bias and the energy resolution are important quantities and directly impact neutrino energy reconstruction for $\nu_e$ interactions.

\par
\begin{figure}[H]
\centering
\begin{subfigure}[b]{0.52\textwidth}
\centering
\includegraphics[width=1.0\textwidth]{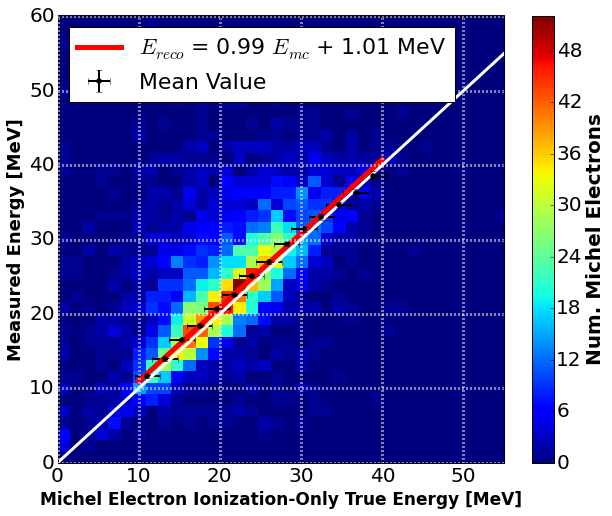}
\caption{}
\end{subfigure}
\begin{subfigure}[b]{0.44\textwidth}
\centering
\includegraphics[width=1.0\textwidth]{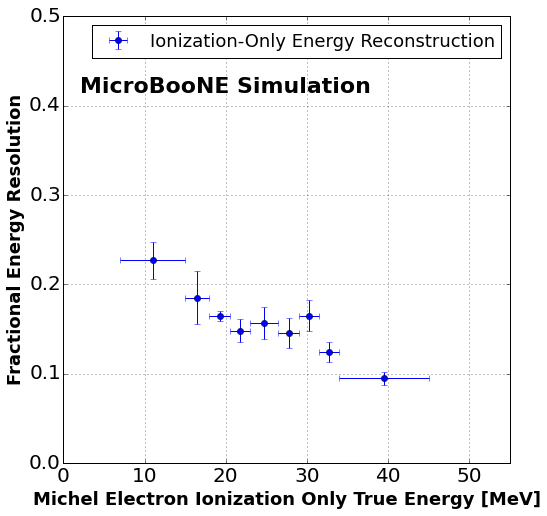}
\caption{}
\end{subfigure}
\caption{(a) reconstructed vs. true ionization-only Michel electron energy deposited. (b) fractional ionization-only energy resolution vs. true ionization-only Michel electron energy.}
\label{fig:eresolution_ioni}
\end{figure}

 Figure~\ref{fig:eresolution_ioni} shows the energy resolution obtained for the ionization-only energy deposited by Michel electrons. This measurement compares the true ionization-only deposited energy to the reconstructed ionization-only clustered energy. The correlation between true and reconstructed energy, quantified by the fitted line in figure~\ref{fig:eresolution_ioni} (a), shows that all ionization energy is correctly accounted for by the reconstruction procedure. The small positive bias (1 MeV) obtained from the fit is due to an over-clustering of charge near the muon decay point where the muon deposits a large amount of charge in its Bragg peak. The fractional energy resolution in figure~\ref{fig:eresolution_ioni} (b) can be interpreted as a measure of the calorimetric resolution, since it compares the measured and true value of the ionization energy that excludes any issue arising from  the energy loss due to radiative photons. The improving resolution as a function of true energy is characteristic of what one would expect for a calorimetric measurement. Ongoing studies in charge-estimation at the hit-reconstruction stage will likely improve the resolution obtained for this kind of measurement.

\par
\begin{figure}[H]
\centering
\begin{subfigure}[b]{0.52\textwidth}
\centering
\includegraphics[width=1.0\textwidth]{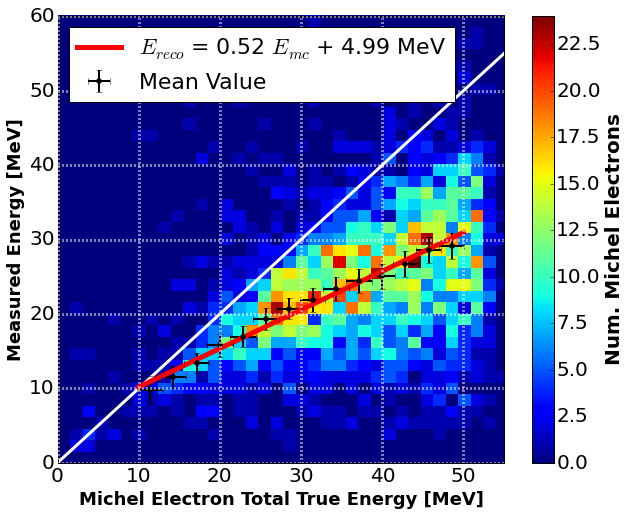}
\caption{}
\end{subfigure}
\begin{subfigure}[b]{0.44\textwidth}
\centering
\includegraphics[width=1.0\textwidth]{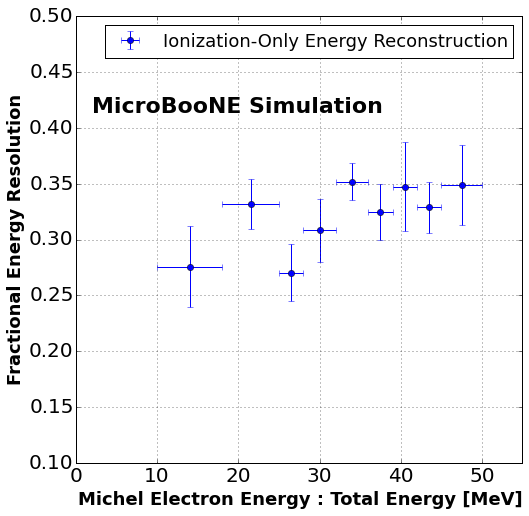}
\caption{}
\end{subfigure}
\caption{(a) reconstructed Michel electron ionization-only energy as a function of the true total Michel electron energy. (b) fractional energy resolution as a function of the true Michel electron energy. Error bars represent the uncertainty on the fit value.}
\label{fig:eresolution_ioni_vs_totl}
\end{figure}

 Figure~\ref{fig:eresolution_ioni_vs_totl} shows the same distributions, but now compares the reconstructed Michel electron energy measured excluding any charge deposited by radiative photons (ionization-only energy measurement) to the true Michel electron energy. Now we see a significant deficit of reconstructed energy, as well as a significant spread in the resolution. Furthermore, we notice that the fraction of collected energy decreases, on average, as the Michel energy increases, and that the distribution itself gets broader. At 50 MeV the reconstructed energy makes up only 60\% of the total true energy implying a significant energy bias. This is consistent with what was shown in figure~\ref{fig:Egammafrac} and once again indicates how an ionization-only energy measurement would not achieve the best energy resolution. 

\par
\begin{figure}[H]
\centering
\begin{subfigure}[b]{0.52\textwidth}
\centering
\includegraphics[width=1.0\textwidth]{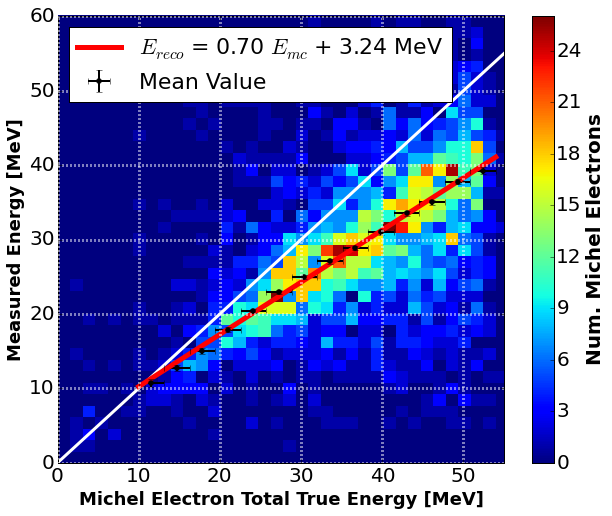}
\caption{}
\end{subfigure}
\begin{subfigure}[b]{0.44\textwidth}
\centering
\includegraphics[width=1.0\textwidth]{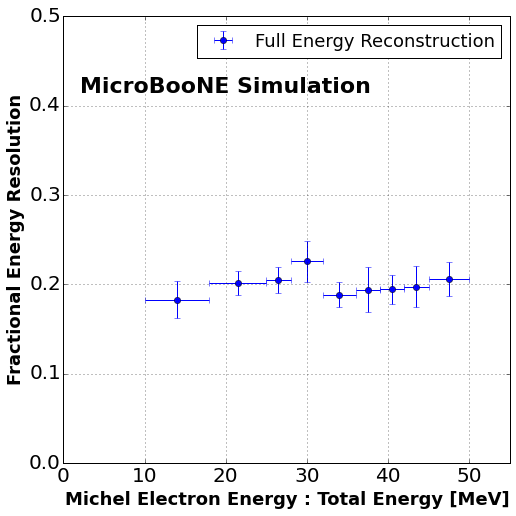}
\caption{}
\end{subfigure}
\caption{(a) reconstructed Michel electron energy as a function of the true Michel electron energy. (b) fractional energy resolution as a function of the true Michel electron energy. The reconstruction procedure to tag radiative photons (described in Sec.~\ref{sec:recoalg}) attempts to collect energy deposited within 80 cm, and \textasciitilde 30 degrees of the muon stopping point in 2D.}
\label{fig:eresolution_totl_totl}
\end{figure}

 Finally, in figure~\ref{fig:eresolution_totl_totl} we show the energy resolution obtained when we include the charge collected through tagged photons. The measured energy is now closer to the true one. Any remaining discrepancy comes from the inefficiency in including radiative photons. For example, at 50 MeV only 76\% of the total energy is recovered, with missing energy attributable to photons under threshold and those mistakenly discarded in an attempt to remove charge from accidental cosmic rays, in addition to energy which escapes the TPC volume or lies beyond the 80 cm and 30 degree cut values used. From a Monte Carlo study of the 2D hit reconstruction employed in this analysis we find that ~15\% of the energy deposited
by photons is under the hit-reconstruction threshold and therefore escapes detection.

\begin{figure}[H]
  \centering
  \includegraphics[width=.6\textwidth]{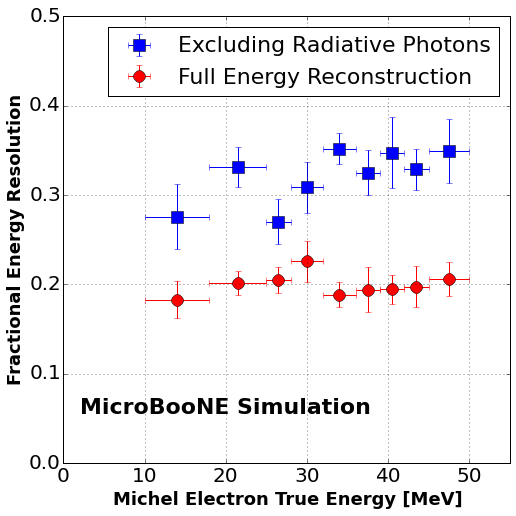}
  \caption{Fractional energy resolution as a function of the true Michel electron energy when using only the reconstructed ionization energy deposited (blue) vs. when including any measured energy deposited by radiative photons (red).}
  \label{fig:eres_comparison}
\end{figure}

\paragraph{}Figure~\ref{fig:eres_comparison} shows the overlay of the resolutions from the ionization-only energy measurement (blue squares) and that including the tagged photons (red circles). Tagging radiative photons improves the energy resolution from over 30\% to 20\%. There is also indication that the improvement is larger at higher energy, where radiative effects are more important. The effect of photon selection and reconstruction still dominates the resolution and bias, so any further improvement will require more efficient photon tagging. This will be challenging, since the presence of the cosmic rays in the detector (approximately at ground level) makes it hard to include all the associated energy without also integrating contributions from nearby cosmic rays.

\section{Reconstructed Michel-Electron Energy Spectrum from Data}
\label{sec:espectrum}

\paragraph{}We finally present the results of the analysis chain described in Sections~\ref{sec:recoalg} and~\ref{sec:ereco} applied to cosmic-ray data recorded by MicroBooNE in ``off-beam'' mode, when the Fermilab accelerator complex was not providing neutrino beams.
\par
\begin{figure}[H]
  \centering
  \includegraphics[width=0.6\textwidth]{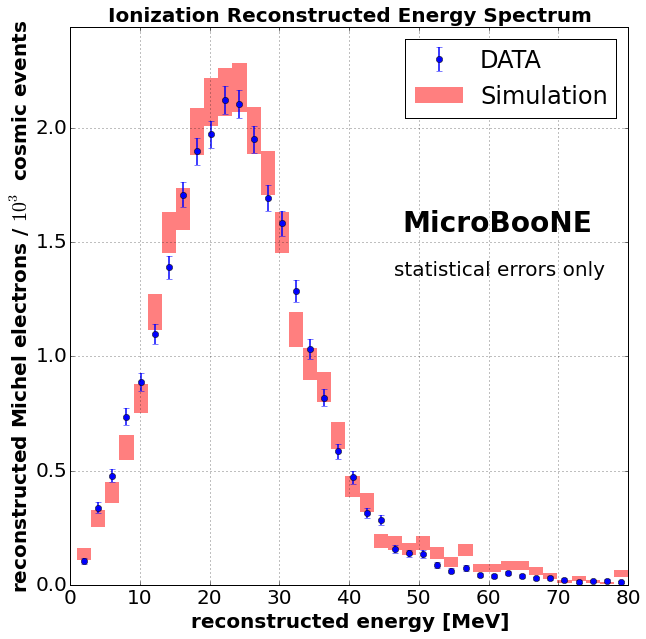}
  \caption{Reconstructed energy spectrum for the measured ionization-only Michel electron energy. The performance of the reconstruction algorithm on data and Monte Carlo events are compared. Error bars represent statistical uncertainties only for both data and simulation.}
  \label{fig:espectrum_ioni}
\end{figure}

\begin{figure}[H]
  \centering
  \includegraphics[width=0.6\textwidth]{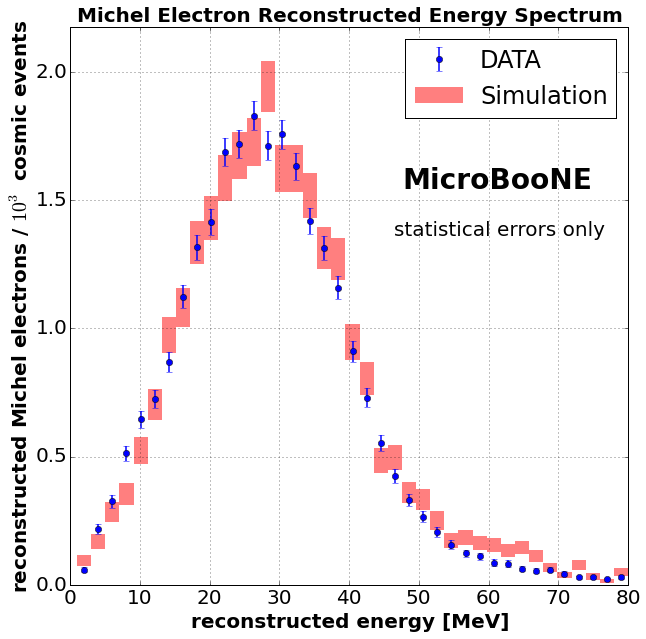}
  \caption{Reconstructed energy spectrum for the full Michel electron energy, including energy associated to tagged photons. The performance of the reconstruction algorithm on data and Monte Carlo events are compared. Error bars represent statistical uncertainties for both data and simulation.}
  \label{fig:espectrum_all}
\end{figure}

 We run the reconstruction chain on a sample of 5.44 $\times 10^5$ triggered events (an event corresponds to a 4.8 ms readout of TPC data) and identify 1.4 $\times 10^4$ candidate Michel electrons. Figure~\ref{fig:espectrum_ioni} shows the energy distribution for the ionization-only energy of tagged electrons. A measurement of the electron ionization spectrum was produced by the ICARUS T600 detector~\cite{bib:icarus_michel}. The two reconstructed spectra show good agreement. Figure~\ref{fig:espectrum_all} shows, for the same events, the reconstructed energy including charge deposited by tagged radiative photons. For both we compare the reconstructed energy spectrum from data to that obtained running over a sample of Monte Carlo simulated cosmic events. No correction for the energy-dependent efficiency is made, which is assumed to be the same for the data and simulation. While the overall reconstruction efficiency is low, it is largely impacted by the spatial orientation of the muon and Michel electron in the 2D collection-plane view and is relatively flat beyond 25 MeV.
\par By performing a $\chi^2$ study of the data and simulation distributions we determine that the energy scale of the two distributions agree to within 3\%. This quantity is compatible with the estimated \textasciitilde5\% systematic uncertainty associated with the calorimetric energy conversion (Eq.~\ref{eq:ADCtoE}). The overall agreement between data and Monte Carlo indicates that our simulation is modeling the physics of muon decay and Michel electron propagation in liquid argon correctly. Furthermore, it allows to expand the conclusions of Sec.~\ref{sec:eres} to the data. For the ionization-only measurement the energy resolution in the range between 10 and 60 MeV is above 30\%, and improves to about 20\% when tagging of radiative photons is included. Including photons shifts the peak of the energy spectrum from \textasciitilde 20 to about 30 MeV. The energy bias measured in Sec.~\ref{sec:eres} changes from 60\% to 76\% of all electron energy recovered at 50 MeV of true electron energy. The decrease in energy-bias and shift in the peak of the energy spectrum are approximately consistent.
\par Both plots show a significant tail which extends beyond the end-point energy of Michel electrons for both data and Monte Carlo. This is due to tagged electrons for which charge from the stopping cosmic-ray muon is incorrectly included in the Michel energy, as well as any possible accidental charge from nearby muons.

\section{Conclusions}
\label{sec:conclusions}

\paragraph{} Low-energy electrons play an important role in many physics studies addressing neutrino oscillations by using the LArTPC detector technology, and are a key ingredient to specific studies, such as the search of neutrino interactions from supernova bursts.  We attempted to highlight the intricate nature of the propagation of such electrons in liquid argon and presented studies illustrating the significant impact of energy lost to radiative photons on the energy resolution.
\par The work presented here shows first results on Michel electron identification with the MicroBooNE TPC employing a simple 2D reconstruction method. A sample of \textasciitilde 14,000 candidate Michel electrons was identified. The reconstructed energy spectrum when only ionization energy deposition is accounted for peaks at \textasciitilde20 MeV and shows a significant energy deficit with respect to the true energy distribution. When we recover energy released by radiative photons, the spectrum's peak shifts to almost 30 MeV. While an improvement, this still falls well below the true energy spectrum cutoff at \textasciitilde 50 MeV. This work shows that we understand the performance of the MicroBooNE LArTPC as evidenced by the Monte Carlo reproducing the observed Michel electron energy spectrum. The reconstructed spectrum is considerably different from the true one, due to the challenge posed by identifying and successfully clustering energy deposited by radiative photons produced by low-energy electrons. We have shown that we can account for some of the missing energy by including tagged radiative photons, hence improving the energy reconstruction. We could improve on this by making full use of the 3D information available from LArTPC detectors. This is an essential step towards reaching the energy resolution required for future precision neutrino oscillation measurements.

\acknowledgments

\paragraph{}This material is based upon work supported by the following:  the U.S. Department of Energy, Office of Science, Offices of High Energy Physics and Nuclear Physics; the U.S. National Science Foundation; the Swiss National Science Foundation; the Science and Technology Facilities Council of the United Kingdom;  and The Royal Society (United Kingdom).   Additional support for the laser  calibration  system  and  cosmic  ray  tagger  was  provided  by  the  Albert  Einstein  Center  for Fundamental Physics. Fermilab is operated by Fermi Research Alliance, LLC under Contract No. DE-AC02-07CH11359 with the United States Department of Energy.


\begin{thebibliography}{9}

\bibitem{bib:SBN}
R. Acciarri et. al., \emph{A Proposal for a Three Detector Short-Baseline Neutrino Oscillation Program in the Fermilab Booster Neutrino Beam}. \href{https://arxiv.org/abs/1503.01520}{arXiv:1503.01520} [hep-ex].

\bibitem{bib:DUNEcdrPhysics}
R. Acciarri et. al, DUNE Collaboration, \emph{Long-Baseline Neutrino Facility (LBNF) and Deep Underground Neutrino Experiment (DUNE) Conceptual Design Report. Volume 2: The Physics Program for DUNE at LBNF}, \href{https://arxiv.org/abs/1512.06148}{arXiv:1512.06148} [hep-ex].

\bibitem{bib:1987a_IMB}
R. Bionta et al., \emph{Observation of a Neutrino Burst in Coincidence with Supernova SN 1987a in the Large Magellanic Cloud}, \emph{Phys. Rev. Lett.}, {\bf 58} (1087), p. 1494.

\bibitem{bib:1987a_Kamiokande}
K. Hirata et al., \emph{Observation of a Neutrino Burst from the Supernova SN 1987a}, \emph{Phys.Rev.Lett.}, {\bf 58} (1987), pp. 1490–1493.


\bibitem{bib:DUNEcdr}
R. Acciarri et. al, \emph{DUNE Collaboration, Long-Baseline Neutrino Facility (LBNF) and Deep Underground Neutrino Experiment (DUNE) Conceptual Design Report}, \href{https://arxiv.org/abs/1601.05471}{arXiv:1601.05471} [hep-ex].

\bibitem{bib:lariat}
J. Paley et al., \emph{LArIAT Collaboration, LArIAT: Liquid Argon In A Testbeam}, \texttt{FERMILAB-PUB-14-268-E},
\href{https://arxiv.org/abs/1406.5560}{arXiv:1406.5560} [hep-ex]. 

\bibitem{bib:LAPD}
M. Adamowski et al., \emph{The Liquid Argon Purity Demonstrator}, \texttt{FERMILAB-PUB-14-059-E},
\href{https://arxiv.org/abs/1403.7236}{arXiv:1403.7236} [physics].

\bibitem{bib:captain}
H. Berns et al., CAPTAIN Collaboration, \emph{The CAPTAIN Detector and Physics Program}, \href{https://arxiv.org/abs/1309.1740}{arXiv:1309.1740} [hep-ex].

\bibitem{bib:35ton}
A. Hahn et al., \emph{The LBNE 35 Ton Prototype Cryostat}, \texttt{FERMILAB-CONF-14-420-PPD}.

\bibitem{bib:argoneut_intro}
C. Anderson et al., ArgoNeuT Collaboration, \emph{The ArgoNeuT Detector in the NuMI Low-Energy beam line at Fermilab}, \emph{JINST}, {\bf 7} (2012) p. 10019, \href{https://arxiv.org/abs/1205.6747}{arXiv:1205.6747} [hep-ex].

\bibitem{bib:icarus_intro}
S. Amerio et al., ICARUS Collaboration, \emph{Design, construction and tests of the ICARUS T600 detector}, \emph{Nucl. Inst. Meth. A}, {\bf527} (2004) pp. 329-410.

\bibitem{bib:icarus_recomb}
S. Amoruso et al., ICARUS Collaboration, \emph{Study of electron recombination in liquid Argon with the ICARUS TPC}, \emph{Nucl. Inst. Meth. A}, {\bf523} (2004) pp. 275-286.

\bibitem{bib:argoneutRecomb}
R. Acciarri et al., ArgoNeuT Collaboration,  \emph{A study of electron recombination using highly ionizing particles in the ArgoNeuT Liquid Argon TPC}, \emph{JINST} {\bf 8} (2013) P08005 \href{https://arxiv.org/abs/1306.1712}{arXiv:1306.1712} [hep-ex].

\bibitem{bib:icarus_lifetime}
M. Antonello et al., ICARUS Collaboration, \emph{Experimental observation of an extremely high electron lifetime with the ICARUS-T600 Lar-TPC}, \emph{JINST}, {\bf 9} (2014). p. 12006, \href{https://arxiv.org/abs/1409.5592}{arXiv:1409.5592} [physics.ins-det].

\bibitem{bib:icarus_pi0}
A. Ankowski et al., ICARUS Collaboration, \emph{Energy reconstruction of electromagnetic showers from $\pi^0$ decays with the ICARUS T600 Liquid Argon TPC}, \emph{Acta Phys. Polon. B} {\bf 41} (2010) pp. 103-125, \href{https://arxiv.org/abs/0812.2373}{arXiv:0812.2373} [physics.ins-det].

\bibitem{bib:argoneut_pi0}
R. Acciarri et al., ArgoNeuT Collaboration, \emph{Measurement of muon neutrino and anti-muon neutrino Neutral Current $\pi^0 \rightarrow \gamma\gamma$ Production in the ArgoNeuT Detector}, \href{https://arxiv.org/abs/1511.00941}{arXiv:1511.00941} [physics.ins-det].

\bibitem{bib:argoneut_dedx}
ArgoNeut Collaboration, R. Acciari et al., \emph{First Observation of Low Energy Electron Neutrinos in a Liquid Argon Time Projection Chamber}, \emph{Phys. Rev. D} {\bf 95}, 072005 \href{https://arxiv.org/abs/1610.04102}{arXiv:1610.04102} [hep-ex].

\bibitem{bib:icarus_michel}
S. Amoruso et al., ICARUS Collaboration, \emph{Measurement of the mu decay spectrum with the ICARUS liquid argon TPC}, \emph{European Physics Journal C} {\bf33} (2004) pp. 233-241. [\href{https://arxiv.org/abs/hep-ex/0311040}{hep-ex/0311040}].

\bibitem{bib:microboone_detectorpaper}
R. Acciarri et al., MicroBooNE Collaboration, \emph{Design and Construction of the MicroBooNE Detector}, \emph{JINST} {\bf 12}, P02017, \href{https://arxiv.org/abs/1612.05824}{arXiv:1612.05824} [hep-ex].

\bibitem{bib:microboone_noise}
R. Acciarri et al., MicroBooNE Collaboration, \emph{Noise Characterization and Filtering in the MicroBooNE Liquid Argon TPC}, \href{https://arxiv.org/abs/1705.07341}{arXiv:1705.07341} [physics.ins-det].

\bibitem{bib:michelNevis}
Marcel Bardon et al., \emph{Measurement of the Momentum Spectrum of Positrons from Muon Decay},
\emph{PRL}, {\bf 14} 449 (1965).

\bibitem{bib:larprop}
Y. Li et al., \emph{Measurement of Longitudinal Electron Diffusion in Liquid Argon}, \emph{Nucl. Inst. Meth. A} {\bf 816} (2016) pp. 160--170, \href{https://arxiv.org/abs/1508.07059}{arXiv:1508.07059} [hep-ex].

\bibitem{bib:bremreview}
H.W. Koch, and J.W. Motz, \emph{Bremsstrahlung Cross-Section Formulas and Related Data}, \emph{Rev. Mod. Phys.} {\bf 31}, {\bf 920} (1959).

\bibitem{bib:radlenPDG}
Particle Data Group, Atomic and nuclear properties of liquid argon (Ar) [{\footnotesize \url{http://pdg.lbl.gov/2012/AtomicNuclearProperties/MUON_ELOSS_TABLES/muonloss_289.pdf}}, retrieved Feb. 20, 2017].

\bibitem{bib:geant4}
 S. Agostinelli et al., GEANT4: A Simulation toolkit, NIM, A506, 250-303, (2003).

\bibitem{bib:NIST_ESTAR}
M.J. Berger, J.S. Coursey, M.A. Zucker, and J. Chang, \emph{ESTAR, PSTAR, and ASTAR: Computer Programs for Calculating Stopping-Power and Range Tables for Electrons, Protons, and Helium Ions} (version 1.2.3), (2005), [\url{http://physics.nist.gov/Star}, retrieved Feb. 24, 2017].

\bibitem{bib:pdg}
 K.A. Olive et al., Particle Data Group, \emph{Chin. Phys. C}, {\bf 38} (2014) 090001.

\bibitem{bib:NIST_XCOM}
 M.J. Berger et al., \emph{XCOM: Photon Cross Section Database (version 1.5)}, (2010),  [\url{http://physics.nist.gov/xcom}, retrieved Feb. 24, 2017].

\bibitem{bib:CRT}
M. Auger et al., \emph{A Novel Cosmic Ray Tagger System for Liquid Argon TPC Neutrino Detectors}, \emph{Nucl. Instruments and Methods A}, {\bf 1}, 2, (2017), \href{https://arxiv.org/abs/1612.04614}{arXiv:1612.04614} [hep-ex].

\bibitem{bib:boundmudecay}
A. Czarnecki et al., \emph{Michel decay spectrum for a muon bound to a nucleus}, \emph{Phys. Rev. D.}, {\bf 90} (2014) no.9, 093002 \href{https://arxiv.org/abs/1406.3575}{arXiv:1406.3575} [hep-ph].

\bibitem{bib:corsika}
D. Heck et al., \emph{CORSIKA: A Monte Carlo Code to Simulate Extensive Air Showers}, \emph{Wissenschaftliche Berichte}, FZKA 6019, (1998).

\bibitem{bib:Wion1}
 M.E. Shibamura et al., \emph{Drift velocities of electrons, saturation characteristics of ionization and W-values for conversion electrons in liquid argon, liquid argon-gas mixtures and liquid xenon}, \emph{Nucl. Instrumentation Meth.}, {\bf 131}, (1975) p249.

\bibitem{bib:Wion2}
M. Miyajima et al., \emph{Average Energy Expended per Ion Pair in Liquid Argon, Liquid Argon-gas Mixtures and Liquid Xenon}, \emph{Phys. Rev. A} {\bf 9} (1974) 1438, and erratum in A10 (1974) 1452.

\end{thebibliography}
\end{document}